%% file: main.tex
\documentclass[cameraready]{Interspeech}

\title{TF-MoE: Time-Frequency Mixture-of-Experts for Efficient Speech Separation}

\author[affiliation={1}, orcid=0009-0004-8699-2916]{Qinzhe}{Hu}
\author[affiliation={1}, orcid=0000-0003-0299-9914]{Chenda}{Li}
\author[affiliation={1}, orcid=0000-0003-4500-3515]{Wangyou}{Zhang}
\author[affiliation={2}, orcid=0009-0008-2599-6752]{Shujie}{Liu}
\author[affiliation={2}, orcid=0000-0003-0384-0219]{Yan}{Lu}
\author[affiliation={1}, orcid=0000-0002-0314-3790]{Yanmin}{Qian}

\address{
    $^1$ Auditory Cognition and Computational Acoustics Lab\\
    Shanghai Jiao Tong University, Shanghai, China \\
    $^2$ Microsoft Research Asia, China
}

\email{lichenda1996@sjtu.edu.cn, yanminqian@sjtu.edu.cn}

\keywords{Speech Separation, Mixture-of-Experts, Computational Efficiency}

\usepackage{comment}

\begin{document}

\maketitle

\begin{abstract}
Recent advances in speech separation (SS) have led to compact front-end models with small parameter sizes, yet their high computational cost remains a major barrier for deployment on edge devices. To address this, we propose TF-MoE, a sparse Mixture-of-Experts (MoE) framework that enhances model capacity with almost no increase in inference cost. Our method introduces dynamic expert specialization in time and frequency dimensions through alternating time-wise and frequency-wise MoE modules, which dynamically select experts per mel band and per time frame, respectively. Built upon a mel-band-splitting Conformer backbone, TF-MoE achieves strong performance on SS tasks under low-compute settings. Experimental results demonstrate that TF-MoE consistently improves separation performance under computation cost constraints, outperforming BSRNN by +3.8 dB SDR on Libri2Mix with comparable 4.1 GMACs/s inference cost. This positions TF-MoE as a promising candidate for edge-device deployment.
\end{abstract}

\input{intro}

\input{methods}

\input{exp}

\section{Conclusion}
\label{sec:conclusion}

We proposed TF-MoE, a highly efficient time-frequency Mixture-of-Experts framework, to overcome the capacity-computation bottleneck in the state-of-the-art speech separation. By integrating sparse expert routing into a dual-path Conformer backbone, our model successfully converts newly added parameters into separation performance almost without increasing inference cost, presenting a highly feasible pathway toward edge-device deployment.


Crucially, our routing analysis reveals that the architecture explicitly decouples acoustic complexity: the experts of the F-module, selected per time frame, dynamically adapt to varying speaker states and overlap events, while the experts of the T-module, selected per mel band, specialize structurally across heterogeneous mel-bands.

\section{Acknowledgments}
This work was supported in part by the China STI 2030--Major Projects under Grant No. 2021ZD0201500, in part by the National Natural Science Foundation of China under Grant No. U25A20409, and in part by the SJTU Med-X (Medicine \& Engineering) Translational Research Grant under Grant No. YG2025LC09.




\section{Generative AI Use Disclosure}
During the preparation of this work, the author(s) used generative AI tools (e.g., ChatGPT, GLM, Gemini) for language polishing and grammar checking to improve the readability of the manuscript. All scientific content, methodology design, and experimental analysis were conducted and verified solely by the authors, who take full responsibility for the final content of this paper.

\bibliographystyle{IEEEtran}
\bibliography{refs}

\end{document}

%% file: intro.tex
\section{Introduction}
\label{sec:intro}

Speech separation (SS) has made great progress with deep learning~\cite{wangSupervisedSpeechSeparation2018a,yuPermutationInvariantTraining2017,luoConvTasNetSurpassingIdeal2019,welkerSpeechEnhancementScoreBased2022,wangTFGridNetMakingTimeFrequency2023,yuEfficientMonauralSpeech2023,zhangURGENTChallengeUniversality2024,koheiurgent2025}, emerging as a critical enabler for a wide range of real-world applications. 
Driven by growing privacy concerns~\cite{tomashenko2024voiceprivacy}, the demand for low-latency interaction~\cite{dubey2024icassp}, and the need for offline capability in network-unavailable environments~\cite{fedorov2020tinylstms,schroter2022low}, there is a surging trend towards deploying SS models under resource-constrained conditions.

In recent academic research, many speech separation (SS) models exhibit highly compact parameter sizes, typically fewer than 10 million parameters~\cite{wangTFGridNetMakingTimeFrequency2023,yuEfficientMonauralSpeech2023,luoDualPathRNNEfficient2020,subakanAttentionAllYou2021,liSkimSkippingMemory2022,xu2024tiger}. 
However, this small parameter footprint is not a true reflection of efficiency.  
These parameter-compact SS models often incur substantial computational costs, reaching tens or even hundreds of giga multiply-accumulate operations per second (GMACs/s)~\cite{wangTFGridNetMakingTimeFrequency2023,yuEfficientMonauralSpeech2023,luoDualPathRNNEfficient2020,subakanAttentionAllYou2021,liSkimSkippingMemory2022,xu2024tiger,kimStackLessRepeat2025}.

This creates a mismatch with the practical constraints of edge deployment. The small model size is generally not a bottleneck, since memory is relatively inexpensive. But high computational cost poses a significant challenge by preventing real-time processing and causing prohibitive power consumption. Recent studies have highlighted this imbalance, revealing that current SS models are effectively under-parameterized when compared to their high computational cost~\cite{kimStackLessRepeat2025,zhangPerformancePlateausComprehensive2024}. For SS models, computational cost plays a more critical role in determining performance than parameter count.
Consequently,  the key to advancing SS for edge deployment lies in correcting this mismatch: increasing model capacity through additional parameters without increasing computational cost.

Recently, many studies have explored various strategies for designing efficient and compact SS models~\cite{yuEfficientMonauralSpeech2023,liSkimSkippingMemory2022,xu2024tiger,liPredictiveSkimContrastive2023,tanCheapNETImprovingLightweight2023,liTFSkiMNetSpeechEnhancement2025}. One of the most straightforward and commonly adopted approaches is to scale down the hidden dimensions of the network. However, excessively reducing the parameter count inevitably leads to a decline in model capacity, which in turn compromises performance~\cite{zhangPerformancePlateausComprehensive2024}.

\begin{figure*}[htb]
  \centering
\centerline{\includegraphics[width=\textwidth]{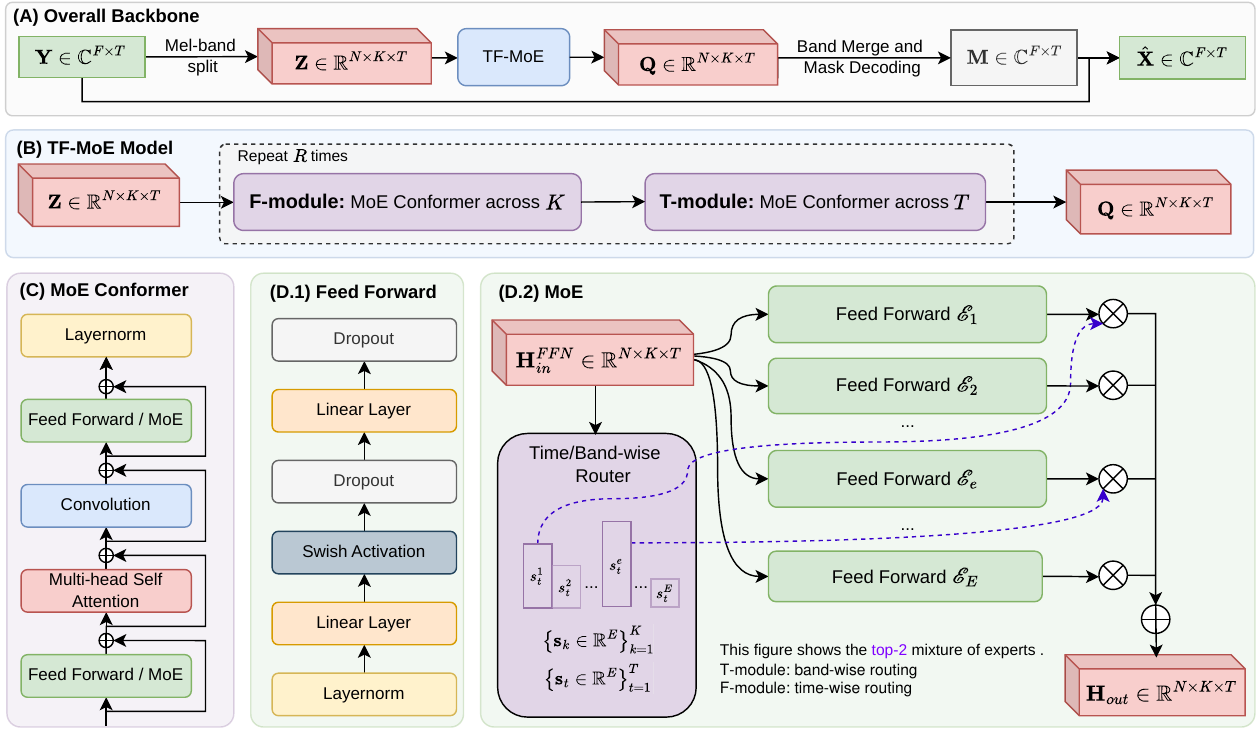}}
\caption{Overview of the proposed TF-MoE framework. (A)~The TF-Conformer backbone; (B)~The proposed TF-MoE model; (C)~The MoE Conformer block, where standard feed-forward modules are replaced by MoE FFNs; (D.1):~The standard feed-forward module;  (D.2)~The MoE feed-forward module with top-$J$ sparse routing.}
\label{fig:tfmoe}
\end{figure*}

Mixture-of-Experts (MoE)~\cite{muComprehensiveSurveyMixtureofExperts2025} models offer a promising solution to this capacity-computation dilemma. An MoE layer consists of multiple expert sub-networks and a gating mechanism that activates only a subset of them per input~\cite{shazeerOutrageouslyLargeNeural2017}. By routing each input to certain experts, MoEs achieve conditional computation, effectively scaling model capacity (\#parameters) without significantly scaling the compute cost. Sparsely-gated MoE architectures have demonstrated orders-of-magnitude larger model capacity at fixed computational cost in other deep learning tasks ~\cite{kumataniBuildingGreatMultilingual2022,duGLaMEfficientScaling2022,daiDeepSeekMoEUltimateExpert2024,zhangCLIPMoEBuildingMixture2025}.

In this paper, we propose \textit{TF-MoE}, a sparse MoE framework that operates in both \textit{time} and \textit{frequency} domain for SS. An overview of the proposed framework is shown in Fig.~\ref{fig:tfmoe}.
It is worth noting that the MoE method has been explored in previous  studies~\cite{chazanSpeechEnhancementUsing2017,wangHandlingTradeOffsSpeech2023}, but these works focused on phoneme-level gating for enhancement~\cite{chazanSpeechEnhancementUsing2017} or applied MoE sparsely along the temporal dimension only~\cite{wangHandlingTradeOffsSpeech2023} for SS.
In contrast, our proposed \textit{TF-MoE} framework introduces a unified \textit{time-frequency expert} structure that performs sub-band-level and frame-level routing across both dimensions. This design enables specialized processing for diverse acoustic patterns, achieving efficient capacity scaling for SS with almost no increase in computational overhead. Our contributions include:

1) We propose a competitive mel-band-splitting Conformer backbone that balances performance and efficiency, 
achieving a +2.5 dB SDR improvement over BSRNN on Libri2Mix at  comparable computational cost.

2) We propose TF-MoE, a sparse Mixture-of-Experts framework that replaces the feed-forward modules in both the time and frequency Conformer blocks with sparsely-gated expert layers, scaling model capacity without increasing computational cost. 
It outperforms Conformer backbone by +1.3 dB SDR on Libri2Mix.

3) Comprehensive ablation studies validate the complementary nature of sparse MoE routing and  Conformer on both temporal and frequency dimensions. Furthermore, visual analysis of the gating policies reveals an explicit structural specialization on different acoustic patterns. 


%% file: methods.tex
\section{Methodology}
\label{sec:methodology}

\subsection{Backbone Model}
\label{ssec:backbone}
We first introduce a new SS model,  TF-Conformer, as our backbone, which is inspired by the band-split RNN (BSRNN)~\cite{yuEfficientMonauralSpeech2023} with two key modifications: (i)~the manually designed sub-band splitting is replaced with mel-scale~\cite{volkmannScaleMeasurementPsychological1937} band splitting, and (ii)~the RNN sequence modeling modules are replaced with Conformer~\cite{gulatiConformerConvolutionaugmentedTransformer2020} blocks.

As shown in Fig.~\ref{fig:tfmoe}.A, the backbone model takes the complex spectrum $\mathbf{Y}\in\mathbb{C}^{F\times T}$ of the mixture signals as input, where $F$ is the number of frequency bins and $T$ denotes the number of time frames.
The mel-band splitting module splits the $F$ frequency bins into $K$ mel-bands and projects the frequency bins in each band into an  $N$-dimensional feature space, resulting in a three-dimensional deep feature representation $\mathbf{Z}\in \mathbb{R}^{N \times K \times T}$.

The core of the model consists of $R$ stacked TF-blocks (Fig.~\ref{fig:tfmoe}(B)), each comprising a frequency Conformer module (\textit{F-module}) that models inter-band dependencies along the $K$ dimension and a temporal Conformer module (\textit{T-module}) that captures temporal dynamics along the $T$ dimension.
Each Conformer module follows the macaron-style architecture~\cite{gulatiConformerConvolutionaugmentedTransformer2020}, consisting of two feed-forward modules (FFN), a multi-head self-attention module, and a convolution module (Fig.~\ref{fig:tfmoe}(C)). 
Crucially, these FFN layers serve as the primary target for our proposed sparse MoE replacement (Sec.~\ref{ssec:moe_ffn}).
After the $R$ TF-blocks, a mask decoding module estimates complex masks $\mathbf{M}\in\mathbb{C}^{F\times T}$ to reconstruct the target speech.

\subsection{Sparse MoE Feed-Forward Module}
\label{ssec:moe_ffn}

To scale model capacity without increasing computational cost, 
we replace the standard FFN  with a sparsely-gated 
Mixture-of-Experts (MoE) FFN.
A standard FFN (Fig.~\ref{fig:tfmoe}(D.1)) consists of two linear projections with a non-linear activation in between:
\begin{align}
    \text{FFN}(\mathbf{x}) = \mathbf{W}_2 \,\sigma(\mathbf{W}_1 
    \mathbf{x} + \mathbf{b}_1) + \mathbf{b}_2,
\end{align}
where $\sigma(\cdot)$ denotes the Swish 
activation~\cite{ramachandran2017searching}.

In the MoE FFN (Fig.~\ref{fig:tfmoe}(D.2)), we maintain $E$ parallel expert networks $\{\mathscr{E}_e\}_{e=1}^{E}$, each structurally identical to the standard FFN but with independent parameters.
The MoE FFN operates on a sequence $\mathbf{X}=[\mathbf{x}_1,\dots,\mathbf{x}_L]\in\mathbb{R}^{L\times N}$, i.e., a sub-band of $L\!=\!T$ frames in the T-module or a frame of $L\!=\!K$ bands in the F-module (Sec.~\ref{ssec:tf_moe}).
A lightweight gating network $\mathcal{G}$, referred to as the router in Fig.~1(D.2), computes a routing distribution over all experts from the mean-pooled descriptor $\bar{\mathbf{x}}=\frac{1}{L}\sum_{l=1}^{L}\mathbf{x}_l$, and only the top-$J$ experts with the highest scores are activated:
\begin{align}
    \label{eq:gate}
    \mathcal{G}(\bar{\mathbf{x}}) &= \text{TopJ}\big(\text{Softmax}
    (\mathbf{W}_g \bar{\mathbf{x}})\big) 
    \rightarrow \{(e_j,\, w_j)\}_{j=1}^{J},
\end{align}
where $\mathbf{W}_g \in \mathbb{R}^{E \times N}$ is the gating projection, and $w_j$ is the normalized weight for the $j$-th selected expert.
The selected experts are shared by all tokens of the sequence, and the MoE FFN output for each token is a weighted sum of their outputs:
\begin{align}
    \label{eq:moe_output}
    \text{MoE-FFN}(\mathbf{x}_l) = \sum_{j=1}^{J} w_j \cdot 
    \mathscr{E}_{e_j}(\mathbf{x}_l),\quad l=1,\dots,L.
\end{align}

When $J \!=\! 1$, every token in the sequence is processed by exactly one expert, matching the expert-computation cost of a standard FFN while increasing FFN parameters by $E$ times.  

To encourage balanced expert utilization and prevent collapse to a few dominant experts, we adopt an auxiliary balance loss~\cite{daiDeepSeekMoEUltimateExpert2024} that encourages uniform expert utilization:
\begin{align}
    \label{eq:total_loss}
    \mathcal{L} = \mathcal{L}_{\text{SS}} + \alpha \,
    \mathcal{L}_{\text{balance}},
\end{align}
where $\alpha$ controls the strength of the balancing regularization.

We now analyze the computational overhead introduced by MoE. For a standard FFN processing a sequence of $L$ tokens $\mathbf{x}_l\in\mathbb{R}^{N}$, the multiply-accumulate operations (MACs) are $L \cdot 2N^2 M$,
where $M\!=\!4$ is the FFN expansion factor.

In the MoE FFN with $E$ experts and top-$J$ routing, the total MACs consist of two parts: gating and expert computation. Since the router acts once per sequence on the pooled descriptor $\bar{\mathbf{x}}$, the gating network requires only $N \cdot E$ MACs per sequence. Each activated expert incurs the same cost as a standard FFN. Thus:
\begin{align}
    \text{MACs}_{\text{MoE}} = \underbrace{N \cdot E}_{\text{gating}} 
    + \underbrace{J \cdot L \cdot 2N^2 M}_{\text{experts}}.
\end{align}
When $J\!=\!1$ (our default setting), the expert computation cost equals that of a single standard FFN, and the gating overhead is negligible:
\begin{align}
    \frac{\text{MACs}_{\text{MoE}}}{\text{MACs}_{\text{FFN}}} 
    = 1 + \frac{E}{2LNM} \approx 1.
\end{align}
For example, with $N=32$, $M=4$, $E=12$, and $L=80$ mel-bands, the gating adds only $\sim$0.06\% to the FFN computation (less than $10^{-3}$ GMACs/s for the entire model). Meanwhile, the corresponding parameter count scales by a factor of $E$ (from $2N^2M$ to $2EN^2M$), achieving the \emph{compute-neutral capacity scaling}.

\subsection{TF-MoE: Dual-Dimension Expert Routing}
\label{ssec:tf_moe}

The MoE FFN described in Sec.~\ref{ssec:moe_ffn} can be applied to the Conformer blocks in either or both of the two modeling dimensions. 
As introduced in Sec.~\ref{ssec:backbone}, the backbone processes the representation $\mathbf{Z}\in\mathbb{R}^{N\times K\times T}$ along the temporal and frequency dimensions via Conformer modules. 
Depending on the target dimension, the routing granularity of the MoE gating differs: 

\textbf{T-MoE.}
In the T-module, $\mathbf{Z}$ is reshaped to $(BK,\,T,\,N)$ so that each of the $K$ sub-bands is processed independently along the $T$ dimension. The MoE gating in Eq.~\eqref{eq:gate} then routes each \emph{sub-band} to a sparse subset of experts, allowing specialized processing for different spectral regions---e.g., dedicating certain experts to low-frequency harmonics and others to high-frequency components.

\textbf{F-MoE.}
In the F-module, $\mathbf{Z}$ is reshaped to $(BT,\,K,\,N)$ so that each of the $T$ time frames is processed independently along the $K$ dimension. The MoE gating routes each \emph{time frame} to a sparse subset of experts, enabling the model to dynamically allocate capacity based on temporal content---e.g., assigning different experts to voiced, unvoiced, and silence segments.

\textbf{TF-MoE.}
By equipping both modules with MoE FFNs, the full 
\textit{TF-MoE} model performs expert routing in both dimensions. 
This dual-dimension design maximizes the capacity scaling: with $E$ experts and top-1 routing in both modules, the model contains approximately $E$ times more FFN parameters than the dense baseline while maintaining the same computational cost.

%% file: exp.tex
\section{Experiments}
\label{sec:exp}

\subsection{Experimental Setup}
\textbf{Datasets \& Configuration:} Experiments are conducted on the Libri2Mix (16kHz, min) dataset from LibriMix \cite{cosentino2020librimix}.
The input spectrogram is extracted using a 32\,ms Hanning window with an 8\,ms shift, and split into $K$=80 mel-scale~\cite{volkmannScaleMeasurementPsychological1937} sub-bands. 
Unless otherwise stated, models use hidden dimension $N=32$ and $R=6$ repeated blocks. For TF-MoE, the default routing uses  top-$J=1$ , and the weight $\mathbf{\alpha}$ for $\mathcal{L}_{\mathrm{balance}}$ is $10^{-3}$.

\noindent\textbf{Training \& Evaluation:}  
AdamW optimizer~\cite{loshchilov2017decoupled} with a cosine annealing scheduler~\cite{loshchilovSGDRStochasticGradient2017} was used for training.  
We adopt the SI-SNR loss~\cite{luoConvTasNetSurpassingIdeal2019} with permutation invariant training (PIT)~\cite{yuPermutationInvariantTraining2017}. 
We report signal-to-distortion ratio (SDR)~\cite{fevotteBSS_EVALToolboxUser2005}, scale-invariant SDR (SI-SDR)~\cite{rouxSDRHalfbakedWell2019}, Perceptual Evaluation of Speech Quality (PESQ)~\cite{rix2001perceptual}, and Short-Time Objective Intelligibility (STOI)~\cite{taalAlgorithmIntelligibilityPrediction2011} to evaluate the quality of separated speech. We also report model parameters (Params),computational cost (MACs/s), and real-time factor (RTF) to evaluate efficiency. RTF is measured on a single-thread laptop CPU by averaging 100 runs on a 2.4-second sample.

\begin{table}[t]
\centering
\caption{Separation results on Libri2Mix 16 kHz.*: Results cited from~\cite{xu2024tiger}.
$^\dagger$:Results cited from~\cite{liSPMambaStatespaceModel2024}, trained and  evaluated on Libri2Mix 8 kHz. }
\label{tab:main_results}
\setlength{\tabcolsep}{3pt}
\renewcommand{\arraystretch}{1}
\resizebox{\columnwidth}{!}{%
\begin{tabular}{lccccccc}
\toprule
Model & Params & MACs/s & RTF & SDR & SI-SDR & STOI & PESQ \\
      & (M)    & (G)    &  & (dB) & (dB)  & (\%) &      \\
\midrule
TF-GridNet$^{*}$ & 14.4 & 323.8& --  & 19.6 & 19.2 & -- & --\\
SPMamba$^\dagger$    & 6.1  & 238.7& --  & 20.4 & 19.9 & -- & -- \\
A-FRCNN-16$^{*}$&6.1 &81.3 & -- & 16.7 & 16.3 & --& -- \\
DualPathRNN$^{*}$& 2.7 & 45.0& --  & 11.6 & 11.3 & -- & -- \\
TDANet Large$^{*}$&2.3& 9.2 & -- & 16.1 & 15.6 & -- & -- \\
Tiger$^{*}$&0.8 & 7.7 & -- & 17.1 & 16.7 & -- & -- \\
Conv-TasNet$^{*}$& 5.6 &7.2& -- & 12.5 & 12.1 & -- & -- \\
SudoRM-RF1.0x$^{*}$  & 2.7 & 4.7 & --   & 13.6 & 13.2 & -- & -- \\

BSRNN              & 2.4 & 4.2& 0.23  & 13.9 & 13.4 & 92.6 & 2.31\\
\midrule
TF-Conformer      & 2.3 & 4.1& 0.45 & 16.4 & 16.0 & 95.4 & 2.63\\
\quad +\textbf{TF-MoE}    & \textbf{4.6} & \textbf{4.1}& 0.47  
& \textbf{17.7} & \textbf{17.2} & \textbf{96.3} & \textbf{2.81}\\
\bottomrule
\end{tabular}%
}
\end{table}

\subsection{Main Results}
\label{ssec:main_results}

Table~\ref{tab:main_results} compares our proposed models against several mainstream models from the literature~\cite{luoConvTasNetSurpassingIdeal2019,wangTFGridNetMakingTimeFrequency2023,yuEfficientMonauralSpeech2023,luoDualPathRNNEfficient2020,liSPMambaStatespaceModel2024,hu2021speech,xu2024tiger,li2022efficient,tzinis2020sudo}. While our models maintain a lightweight footprint of approximately 4 GMACs/s, existing SOTA models like TF-GridNet~\cite{wangTFGridNetMakingTimeFrequency2023} and SPMamba~\cite{liSPMambaStatespaceModel2024} incur a prohibitively high computational cost—often exceeding 200 GMACs/s—rendering them impractical for real-time edge deployment.

In the computationally constrained and practical regime, our proposed TF-MoE exhibits dominant efficiency. Operating at a highly lightweight footprint of just 4.1 GMACs/s, TF-MoE achieves a remarkable 17.7\,dB SDR, largely outperforming a wide range of popular baselines that consume substantially more computation. For instance, TF-MoE surpasses A-FRCNN-16~\cite{hu2021speech} by \textbf{+1.0\,dB} SDR while consuming nearly $\mathbf{20\times}$ fewer MACs. It also eclipses recent efficient designs like Tiger~\cite{xu2024tiger} (7.7G, 17.1\,dB) and TDANet Large~\cite{li2022efficient} (9.2G, 16.1\,dB) in both separation quality and computational economy. The glaring performance gap between classical models like DualPathRNN~\cite{luoDualPathRNNEfficient2020}(45.0G, 11.6\,dB) or Conv-TasNet~\cite{luoConvTasNetSurpassingIdeal2019} (7.2G, 12.5\,dB) and our framework vividly illustrates the capacity-computation mismatch discussed in Sec.~\ref{sec:intro}.

Within our model family, the TF-Conformer backbone already outperforms BSRNN~\cite{yuEfficientMonauralSpeech2023} by \textbf{+2.5\,dB} SDR at comparable cost, validating the mel-scale band-split Conformer design. Building upon this backbone, TF-MoE ($E\!=\!12$, $J\!=\!1$) achieves a further \textbf{+1.3\,dB} improvement while keeping MACs/s almost unchanged. The additional parameters introduced by sparse expert layers are effectively converted into performance gains with almost no computational overhead, confirming the effectiveness of compute-neutral capacity scaling via MoE.

\subsection{Ablation Studies}
\label{ssec:ablation}
We first ablate the architectural components with different backbone configurations, and then analyze the learned routing to interpret why the proposed configuration performs best.

\noindent\textbf{Ablation on Architectural Components.}
Table~\ref{tab:ablation_dim} presents a systematic ablation by progressively replacing the advanced module with simpler components: MoE~$\rightarrow$~Conformer (removing mixture-of-experts ) and Conformer~$\rightarrow$~RNN.

\begin{table}[t]
\centering
\caption{Ablation on T/F-module.  All MoE variants use $E\!=\!6$, $J\!=\!1$. $^\ddagger$: hidden dimension reduced from 32 to 30 to match the computational cost of MoE variants. All models have almost the same GMACs/s=4.1.}
\label{tab:ablation_dim}
\setlength{\tabcolsep}{3pt}
\renewcommand{\arraystretch}{1}
\small
\resizebox{\linewidth}{!}{%
\begin{tabular}{lcc cc cc}
\toprule
Model & T-module & F-module & Params  & SDR & SI-SDR \\
         &          &     & (M)    & (dB) & (dB)  \\
\midrule
BSRNN & RNN & RNN & 2.4 & 13.9 & 13.4  \\
\multicolumn{6}{l}{\textit{(a) TF-Conformer ablation}} \\
TF-conformer&Conformer      & Conformer      & 2.3 &  16.4 & 16.0 \\
\quad - w/o T-Conformer$^\ddagger$ &RNN  &Conformer      & 2.1 & 16.1 & 15.6 \\
\quad - w/o F-Conformer$^\ddagger$ & Conformer      & RNN & 2.1  & 16.1 & 15.6 \\
\midrule
\multicolumn{6}{l}{\textit{(b) TF-MoE ablation}} \\
TF-MoE ($E=6$, $J=1$) & MoE        & MoE        & 3.4  & \textbf{17.2} & \textbf{16.8} \\
\quad - w/o T-MoE &Conformer  & MoE        & 2.9  & 16.5 & 16.0 \\
\quad \quad - w/o T-Conformer$^\ddagger$ & RNN        & MoE        & 2.8  & 16.3 & 15.8 \\
\quad - w/o F-MoE & MoE        & Conformer  & 2.9  & 17.1 & 16.7 \\
\quad \quad - w/o F-Conformer$^\ddagger$ & MoE        & RNN        & 2.8  & 16.9 & 16.4 \\

\bottomrule
\end{tabular}%
}
\end{table}

The ablation results establish a clear performance hierarchy: RNN $<$ Conformer $<$ MoE. Specifically, replacing either the T or F-module Conformer with an RNN leads to a 0.3\,dB SDR degradation, confirming that both dimensions contribute significantly to the backbone's modeling capability. 
Furthermore, equipping either T or F module with MoE yields  improvements over the Conformer baseline, with the full TF-MoE model achieving the best performance. This validates that the mixture-of-experts mechanism effectively enhances the model's capacity to handle complex acoustic scenarios.

\begin{table}[t]
\centering
\caption{Ablation on expert number $E$ with top-1 routing 
($J\!=\!1$). All models have almost the same GMACs/s=4.1.}
\label{tab:ablation_E}
\setlength{\tabcolsep}{2pt}
\small
\renewcommand{\arraystretch}{0.8}
\begin{tabular}{ccccccc}
\toprule
$E$ & Params (M) & RTF & SDR(dB) & SI-SDR(dB) & STOI(\%) & PESQ \\
\midrule
3   & 2.8 &0.46 & 16.5 & 16.1 & 95.5 & 2.65 \\
6   & 3.4 &0.47 & 17.2 & 16.8& 95.9 & 2.75 \\
12  & 4.6 & 0.47  & \textbf{17.7} & \textbf{17.2}& \textbf{96.3} & \textbf{2.81} \\
24  & 7.0 & 0.49  & 16.6 & 16.1 & 95.6 & 2.66 \\
\bottomrule
\end{tabular}
\end{table}

\noindent\textbf{Expert Capacity and Routing Interpretability.} Table~\ref{tab:ablation_E} investigates the effect of the expert number $E$. Increasing $E$ from 3 to 12 progressively improves performance, as a larger expert pool provides better specialization for diverse acoustic patterns. However, further increasing $E$ to 24 leads to a  performance drop ($-$1.1\,dB SDR compared to $E\!=\!12$). 
This degradation may be caused by the increased difficulty of learning routing policies as the number of experts increases.

\begin{figure}[htb]
  \centering
\centerline{\includegraphics[width=\linewidth]{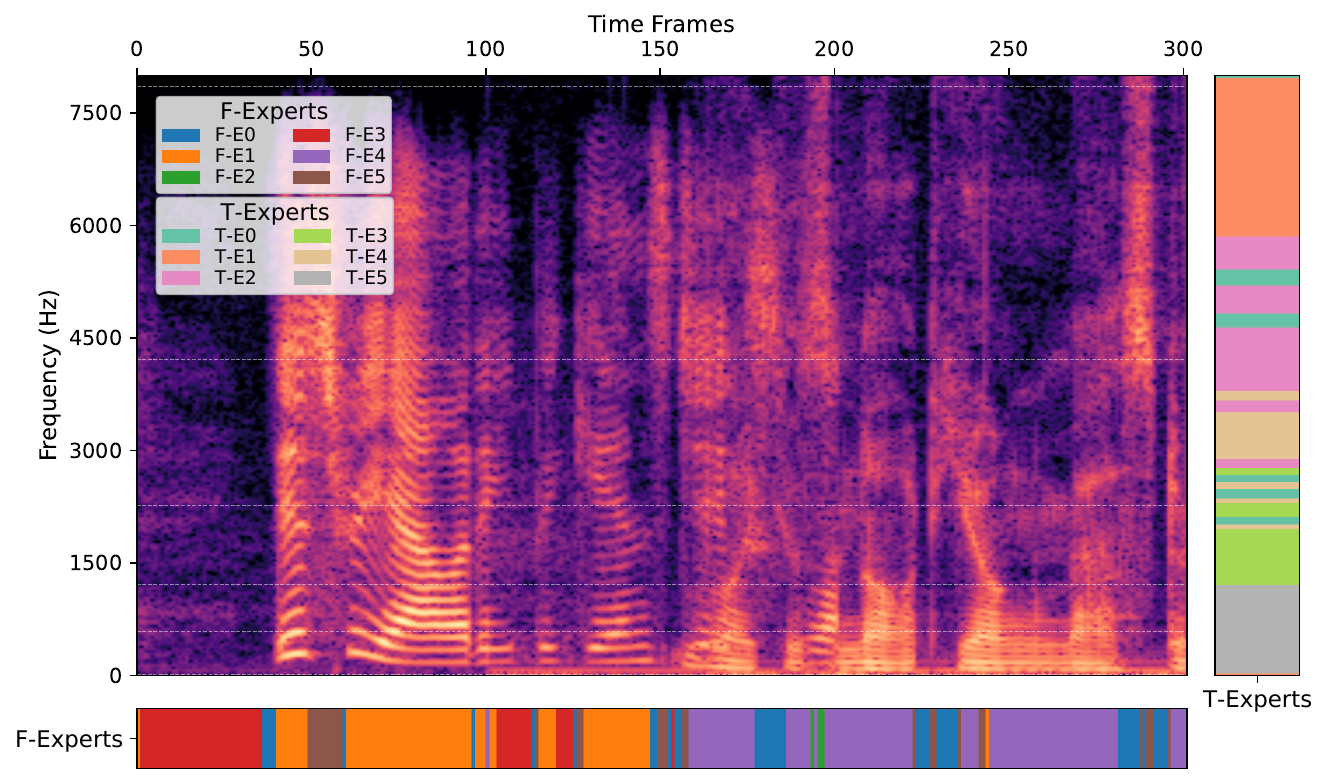}}
\caption{Visualization of the expert routing in the first Feed Forward module of the fifth TF-MoE block. The top spectrogram displays the time-frequency representation of the input signal comprising female-only (frames 0-100), mixed (100-200), and male-only (200-300) segments. The lower bar shows the frame-wise routing decisions of the F-module (one expert per time frame) along the time axis, while the right bar shows the band-wise routing decisions of the T-module (one expert per mel band) along the mel-band axis. Different colors represent distinct experts.}
\label{fig:tfmoe2}
\end{figure}

To understand  how experts utilize the capacity, we visualize the routing decisions in Figure~\ref{fig:tfmoe2}. The visualization reveals that experts have learned distinct specializations corresponding to specific acoustic characteristics:
1) \textbf{Frequency-Specialized T-MoE Experts:} As shown in the right bar chart, the band-wise routing of the T-module exhibits a clear frequency-dependent pattern. Different experts are consistently activated for different mel bands (e.g., low, mid, and high frequencies), indicating that the T-MoE experts have specialized in processing distinct spectral patterns.
2) \textbf{Speaker/Pattern-Sensitive F-MoE Experts:} The lower bar chart shows that the frame-wise routing of the F-module varies significantly across time frames. Distinct experts are activated during different temporal segments, which often correspond to different speakers or varying speaking patterns (e.g., male, female voices, or silence). 

The fact that a finite set of experts can align well with distinct frequency bands and speaker patterns explains why moderate $E$ (e.g., 12) is sufficient, while an excessive number of experts leads to training difficulties and performance degradation.